\documentclass[aps,twocolumn]{revtex4}
\usepackage{amsfonts}
\usepackage{amsmath}

\newcommand{\ket}[1]{\left\vert #1 \right\rangle}
\newcommand{\bra}[1]{\left\langle #1 \right\vert}
\newcommand{\beq}{\begin{equation}}
\newcommand{\eeq}{\end{equation}}

\begin{document}

\title{Unbiased non-orthogonal bases for tomographic reconstruction}
\author{Isabel Sainz$^1$, Luis Roa$^2$, and Andrei B. Klimov$^1$}
\affiliation{$^1$Departamento de F\'isica, Universidad de Guadalajara, Revoluci\'on 1500,
Guadalajara, Jalisco, 44420, M\'exico.\\
$^{2}$Center of Quantum Optics and Quantum Information, Center for Optics
and Photonics, Departamento de F\'{\i}sica, Universidad de Concepci\'{o}n,
Casilla-160C, Concepci\'{o}n, Chile.}

\begin{abstract}
We have developed a general method for constructing a set of
non-orthogonal bases with equal separations between all different basis' states in prime dimensions.It results that the corresponding bi-orthogonal counterparts are pairwise unbiased with the components of the original bases. Using these bases we derive an explicit expression for the optimal
tomography in non-orthogonal bases. Special two dimensional case is analyzed separately.
\end{abstract}

\maketitle

\section{Introduction}

The complementarity between two observables, $A$ and $B$, means that if a
state of a quantum system is measured in the basis of eigenstates
corresponding to the observable $A$, a subsequent measurement, in the basis
of eigenstates of the observable $B$, produces no new information about the
initial quantum state. In other words, the outcomes of one projective
measurement are independent of all the other (projective) measurements. In
this sense, these two bases are unbiased. Two bases, $\left\vert
a;A\right\rangle $ and $\left\vert b;B\right\rangle $ in a $d-$dimensional
Hilbert space are called unbiased if the absolute value of the scalar
product between any two components of different bases is a constant $%
|\langle a;A|b;B\rangle |=1/\sqrt{d}$.

It is well know that if the dimension of the Hilbert space is a power of a
prime number, it is always possible to construct $d+1$ orthogonal mutually
unbiased bases (MUB) \cite{MUB1}. In this case, the $d(d+1)$ mutually
unbiased operators $P_{n}^{s}=|\psi _{n}^{s}\rangle \langle \psi _{n}^{s}|$
define a complete set of projection measurements. The set of $d(d+1)$
projectors decomposes the identity (up to a constant) and the measured
probabilities $p_{sn}=Tr(P_{n}^{s}\hat{\rho}),\label{p}$ completely
determine the $\hat{\rho}$ density operator of the system, which allows to
develop the optimal quantum tomographic procedure \cite{MUBTom}. That is,
when the density oparator is reconstructed in terms of the projectors $
P_{n}^{s}$, each coefficient linearly depends only on the corresponding
probability $p_{sn}$. This results in a great advantage with respect to any
other quantum state tomography schemes, since the error associated with the
measurement of a single projector $P_{n}^{s}$ does not propagate.

The standard tomography methods are usually related to orthogonal
measurements; $Tr(P_{n}^{s}P_{m}^{s})=\delta _{nm}$. A protocol of quantum
state reconstruction using non-orthogonal bases has been applied to a
particular case when only limited access to the full state space is granted
in \cite{non ort tom}. A similar procedure was discussed in the context of
tomographic quantum cryptography \cite{non ort cript}. Also, measurements in
non-orthogonal bases have been used for quantum tomography of photon pairs
with entangled orbital angular momentum \cite{NOtom1}. None of the above
procedures are optimal, in the sense that the elements of the reconstructed
density operator are expressed as some linear combinations of measured
probabilities (more specifically, systems of linear equations should be
solved to determine the density matrix elements).

Among different types of non-orthogonal bases there is a particularly
interesting class of equally separated bases, $\{\left\vert n\right\rangle
,n=1,...,d\}$, such that the overlap between any pair of different states $m\neq n$ is the same, $|\langle m|n\rangle |=\lambda $.
Equally separated bases have been investigated in the context of unambiguous state discrimination \cite{Lucho}.
A particular case of such bases belongs to the class of the so-called symmetric states \cite{chefles}, which are generated by applying integer powers of a single
unitary operator onto any one of the states.
An advantage of the equally
separated bases consists in that the corresponding bi-orthogonal basis \cite
{biort}, $\{\left\vert \tilde{m}\right\rangle \}$ such that $\langle \tilde{m
}|n\rangle =\delta _{mn}$, can be easily constructed. This opens the
possibility to define an explicit optimal tomographic procedure along the
lines of the general approach discussed in \cite{samsonov}.

In this article we generalize the concept of mutually unbiased orthogoal
bases to the bi-orthogonal case. In particular, when the dimension of the
Hilbert space is a prime number $p$,\ we find $p$ equally separated bases
which are unbiased with the set of their bi-orthogonal counterparts. As an
important result we show that using these $p$ bases and a single orthonormal
basis it is possible to reconstruct the density matrix in the optimal way
even when the measurements are performed mostly in the non-orthogonal bases.

\section{Bi-orthogonal mutually unbiased bases}

When the dimension of the Hilbert space is a prime number $p$ we can follow
the standard construction valid for the orthogonal case (see e.g. \cite{MUB
constr}) and obtain $p$ non-orthogonal equally separated bases, such that
their bi-orthogonal counterparts are pairwise unbiased with the original set
of bases.

Let us consider a linearly independent and non-orthogonal set of normalized
states in a $p$-dimensional Hilbert space, $\{\left\vert \psi
_{n}\right\rangle ,n=0,...,p-1\}$, such that the scalar product between any
two different states of the set is a real constant $\lambda $,
\begin{equation}
\langle \psi _{m}|\psi _{n}\rangle =(1-\lambda )\delta _{nm}+\lambda .
\label{basis}
\end{equation}%
The corresponding bi-orthogonal basis $\{\left\vert \phi _{n}\right\rangle
,n=0,\ldots ,p-1\}$, where $\langle \phi _{m}|\psi _{n}\rangle =\delta _{nm}/%
\sqrt{\mu }$, can be expressed in terms of the original basis as,
\begin{equation}
|\phi _{n}\rangle =\sqrt{\mu }|\psi _{n}\rangle +\frac{\nu }{\sqrt{\mu }}
\sum_{\substack{ m=0  \\ m\neq n}}^{p-1}|\psi _{m}\rangle ,  \label{biort1}
\end{equation}%
where
\begin{eqnarray*}
\mu &=&\frac{1+(p-2)\lambda }{(1-\lambda )(1+(p-1)\lambda )},\\ \nu 
&=&-\frac{%
\lambda }{(1-\lambda )(1+(p-1)\lambda )}.
\end{eqnarray*}%
The basis (\ref{biort1}) is normalized, $\langle \phi _{m}|\phi _{n}\rangle
=(1-\eta )\delta _{nm}+\eta $ and equally separated, being $\eta =\nu /\mu $
the separation between the elements of the basis.

The set $\{|\psi _{n}\rangle ,n=0,...p-1\}$ can be considered the
eigenstates of the non-unitary cyclic ($Z^{p}=\hat{\mathbf{I}}$) operator $%
Z|\psi _{n}\rangle =\omega ^{n}|\psi _{n}\rangle $, defined as,
\begin{equation}
Z=\sqrt{\mu}\sum_{m=0}^{p-1}\omega ^{m}|\psi _{m}\rangle \langle \phi _{m}|,
\label{Z}
\end{equation}%
where $\omega =e^{2\pi i/p}$, while the bi-orthogonal basis is formed by
eigenstates of its Hermitian conjugate \cite{biort}, $Z^{\dagger }|\phi
_{n}\rangle =\omega ^{-n}|\phi _{n}\rangle $.

Now, let us introduce the operator
\begin{equation}
X=\sqrt{\mu}\sum_{m=0}^{p-1}|\psi _{m+1}\rangle \langle \phi _{m}|,
\label{X}
\end{equation}
which forms a dual pair\emph{\ }with the operator $Z$, i.e. $ZX=\omega XZ$,
so that the operators $Z$ and $X$ can be considered as generators of the
generalized Pauli group. The operator (\ref{X}) acts as a shift operator,
that is, $X|\psi _{n}\rangle =|\psi _{n+1}\rangle $, so that $|\psi
_{n}\rangle =X^{n}|\psi _{0}\rangle $, and similarly $|\phi _{n}\rangle
=X^{n}|\phi _{0}\rangle $. This property implies that both sets $\{|\psi
_{n}\rangle ,n=0,...,p-1\}$ and $\{|\phi _{n}\rangle ,n=0,...p-1\}$ belong
to the class of symmetric states \cite{chefles}. It is worth noting that
this operator is cyclic, $X^{p}=\hat{\mathbf{I}}$, and unitary, $XX^{\dag
}=X^{\dag }X=\hat{\mathbf{I}}$, so that the basis composed by its
eigenstates is orthogonal.

In close analogy with the orthogonal case, we can find another $p-1$ bases
whose elements correspond to the eigenstates of the monomials $Z^{s}X$, $%
Z^{s}X|\psi _{n}^{s}\rangle =\omega ^{-n}e^{i\phi _{s}}|\psi _{n}^{s}\rangle
$, $s=1,...,p-1,$ where $e^{i\phi _{s}}=i$ for $p=2$, and $e^{i\phi
_{s}}=\omega ^{2^{-1}s}$ for $p>2$. Note that here $2^{-1}$ means the
inverse of 2 $\mod p$. Explicitly, the components of the $s$-th basis are
given by
\begin{equation}
\left\vert \psi _{n}^{s}\right\rangle =\frac{1}{\sqrt{p}}\sum_{m=0}^{p-1}%
\omega ^{2^{-1}sm^{2}+nm}|\psi _{m}\rangle ,\quad s=1,..,p-1,  \label{psi_s}
\end{equation}%
for $p>2$, and $|\psi _{n}^{1}\rangle =\left( \left\vert \psi
_{0}\right\rangle \pm i\left\vert \psi _{1}\right\rangle \right) /\sqrt{2}$
for $p=2$, where from now on the original basis, composed by eigenstates of $%
Z$, will be labelled as the $0$-th basis, $\{\left\vert \psi
_{n}^{0}\right\rangle \equiv \left\vert \psi _{n}\right\rangle \}$.

It can be shown (using the properties of Gauss sums) that the elements of
each basis (\ref{psi_s}) are equally separated with the same absolute value
of the scalar product as the original basis, i.e. $\left\vert \langle \psi
_{n}^{s}\left\vert \psi _{m}^{s}\right\rangle \right\vert =(1-\lambda
)\delta _{nm}+\lambda $ ; the basis elements however are not of the
symmetric type \cite{chefles}. Unlike to the orthogonal case, these bases
are not mutually unbiased, which means that $\left\vert \left\langle \psi
_{n}^{s}|\psi _{m}^{t}\right\rangle \right\vert $ is not a constant for $%
s\neq t$.

Similarly, a corresponding set of $p$ bi-orthogonal bases, $\left\langle
\psi _{n}^{s}|\phi _{m}^{s}\right\rangle =\delta _{nm}/\sqrt{\mu }$, can be
constructed using the basis (\ref{biort1}),
\begin{equation}
\left\vert \phi _{n}^{s}\right\rangle =\frac{1}{\sqrt{p}}\sum_{m=0}^{p-1}%
\omega ^{2^{-1}sm^{2}+nm}\left\vert \phi _{m}\right\rangle ,\quad
s=1,...,p-1,  \label{bi psi s}
\end{equation}%
for $p>2$, while for $p=2$, the bi-orthogonal basis is $\left\vert \phi
_{n}^{1}\right\rangle =(\left\vert \phi _{0}\right\rangle \pm i\left\vert
\phi _{1}\right\rangle )/\sqrt{2}$. The elements of the bases (\ref{bi psi s}%
) are equally separated with the same absolute value as the basis (\ref%
{biort1}). It is clear that the bases (\ref{bi psi s}) are eigenstates of
the monomials $Z^{\dagger s}X$.

Using (\ref{psi_s}), (\ref{bi psi s}) and the definition of bi-orthogonal
bases, one can prove that the set of bases $\{\psi _{n}^{s}\}$ are mutually
unbiased with their corresponding bi-orthogonal counterparts $\{\left\vert
\phi _{m}^{t}\right\rangle \}$, i.e.
\begin{equation}
\left\vert \left\langle \phi _{m}^{t}|\psi _{n}^{s}\right\rangle \right\vert
^{2}=\frac{\delta _{st}\delta _{nm}}{\mu }+\frac{(1-\delta _{st})}{\mu p},
\label{unibiasedness}
\end{equation}%
for all $s,t=0,\ldots ,p-1$ and $n,m=0,\ldots p-1$, where $\{\left\vert \phi
_{n}^{0}\right\rangle \equiv \left\vert \phi _{n}\right\rangle \}$.

\section{Quantum tomography}

We will use the unbiasedness relation (\ref{unibiasedness}) between the
bases $\{\left\vert \psi _{n}^{s}\right\rangle \}$ and $\{\left\vert \phi
_{m}^{t}\right\rangle \}$ for optimal reconstruction of a density matrix $%
\hat{\rho}$ in the $p$-dimensional Hilbert space.

The main idea consists in expanding the density matrix on the projectors $%
\left\vert \phi _{n}^{s}\right\rangle \left\langle \phi _{n}^{s}\right\vert $%
, while the measurements are accomplished in their bi-orthogonal bases $%
\{\left\vert \psi _{n}^{s}\right\rangle \}$. Note that
the set of projectors $\{\ket{\phi_n^s}\bra{\phi_n^s}\}$ and $\{\ket{\psi_n^s}\bra{\psi_n^s}\}$ for $n=0,\ldots p-1,s=0,\ldots p-1$ do not decompose the identity, and thus do not satisfy the condition to form a positive operator valued measure (POVM). To correct
this problem we introduce the (orthonormal) basis of eigenstates of the
unitary operator $X$,
\begin{eqnarray}
\left\vert \psi _{0}^{x}\right\rangle &=&\frac{1}{\sqrt{p(1+(p-1)\lambda )}}%
\sum_{m=0}^{p-1}\left\vert \psi _{m}^{0}\right\rangle ,  \label{psi_00} \\
\left\vert \psi _{n}^{x}\right\rangle &=&\frac{1}{\sqrt{p(1-\lambda )}}%
\sum_{m=0}^{p-1}\omega ^{nm}|\psi _{m}^{0}\rangle ,  \label{psi_0n}
\end{eqnarray}%
where $n=1,\ldots ,p-1$. Observe, that the inversion of the above
expressions solve the problem of defining a non-orthonormal basis with the
overlap function as in Eq.(\ref{basis}) in terms of an orthonormal basis.

Using Eqs.(\ref{psi_00})-(\ref{psi_0n}) we can expand the identity operator
using the non-orthogonal projectors corresponding to the basis $\{\left\vert
\psi _{n}^{s}\right\rangle \}$ as follows,
\begin{eqnarray}
\hat{\mathbf{I}} &=&\mu \sum_{n=0}^{p-1}\left\vert \psi
_{n}^{s}\right\rangle \left\langle \psi _{n}^{s}\right\vert -\nu (1-\lambda
)\sum_{n=1}^{p-1}\left\vert \psi _{n}^{x}\right\rangle \left\langle \psi
_{n}^{x}\right\vert  \notag \\
&&+\nu (p-1)(1+(p-1)\lambda )\left\vert \psi _{0}^{x}\right\rangle
\left\langle \psi _{0}^{x}\right\vert ,  \label{identity1}
\end{eqnarray}%
for $s=0,\ldots ,p-1$, and in a similar way in terms of the corresponding
normalized bi-orthogonal bases $\{\left\vert \phi _{n}^{s}\right\rangle \}$
the identity takes the form
\begin{eqnarray}
\hat{\mathbf{I}} &=&\mu \sum_{n=0}^{p-1}\left\vert \phi
_{n}^{s}\right\rangle \left\langle \phi _{n}^{s}\right\vert -\lambda (\mu
-\nu )\sum_{n=1}^{p-1}\left\vert \psi _{n}^{x}\right\rangle \left\langle
\psi _{n}^{x}\right\vert  \notag \\
&&+\lambda (p-1)\left( \mu +(p-1)\nu \right) \left\vert \psi
_{0}^{x}\right\rangle \left\langle \psi _{0}^{x}\right\vert .  \label{I s}
\end{eqnarray}

Clearly, the elements of the orthogonal basis (\ref{psi_00})-(\ref{psi_0n})
decompose the identity, $\sum_{n=0}^{p-1}\left\vert \psi
_{n}^{x}\right\rangle \left\langle \psi _{n}^{x}\right\vert =\hat{\mathbf{I}}
$, and thus $\sum_{n=0}^{p-1}$ $p_{xn}=1$, where $p_{xn}=\left\langle \psi
_{n}^{x}\right\vert \hat{\rho}\left\vert \psi _{n}^{x}\right\rangle $ are
the measured probabilities in this basis. For the non-orthogonal bases, this
is no longer true, from Eq. (\ref{identity1}) it is easy to see that for any
$s=0,\ldots ,p-1$, the following relation is satisfied
\begin{equation}
\sum_{n=0}^{p-1}p_{sn}=(1-\lambda)+p\lambda p_{x0},  \label{sum NO}
\end{equation}
where $p_{sn}=\left\langle \psi _{n}^{s}\right\vert \hat{\rho}\left\vert
\psi _{n}^{s}\right\rangle $ are the measured probabilities in the bases $%
\{\left\vert \psi _{n}^{s}\right\rangle \}$ for all $s=0,\ldots, p-1$.

The relations (\ref{I s}) and (\ref{sum NO}) allow the expansion of the
density matrix as a sum of projectors on $p$ non-orthogonal equidistant
bases (\ref{psi_s}) and the orthogonal basis (\ref{psi_00})-(\ref{psi_0n}),
as%
\begin{eqnarray}
\hat{\rho} &=&\mu \sum_{s=0}^{p-1}\sum_{n=0}^{p-1}p_{sn}\left\vert \phi
_{n}^{s}\right\rangle \left\langle \phi _{n}^{s}\right\vert  \label{rho_R} \\
&&+\frac{1-\lambda }{1+(p-1)\lambda }\left( p_{x0}-1\right) \left\vert \psi
_{0}^{x}\right\rangle \left\langle \psi _{0}^{x}\right\vert  \notag \\
&&+\sum_{n=1}^{p-1}\left( p_{xn}-p_{x0}p\frac{\lambda }{1-\lambda }-1\right)
\left\vert \psi _{n}^{x}\right\rangle \left\langle \psi _{n}^{x}\right\vert ,
\notag
\end{eqnarray}%
where the expansion coefficients are linear combinations of the
probabilities measured in the corresponding bases $\{\left\vert \psi_n^s
\right\rangle\}$, $s=x,0,\ldots,p-1$. As in the orthonormal case, each
measurement determines a single element of the density matrix, providing in
this sense the optimal reconstruction scheme. In the limit $\lambda
\rightarrow 0$, all the bases $\{\left\vert \psi _{n}^{s}\right\rangle \}$
become orthonormal, each basis coincides with its corresponding
bi-orthonormal counterpart Eq.(\ref{biort1}) and acquire the standard form
\cite{MUB constr}, so that the usual reconstruction expression $\hat{\rho}%
=\sum_{s=0}^{p}\sum_{n=0}^{p-1}p_{sn}\left\vert \psi _{n}^{s}\right\rangle
\left\langle \psi _{n}^{s}\right\vert -\hat{\mathbf{I}} $ is automatically
recovered.

It was argued in \cite{non ort tom} that measuring in non-orthogonal bases
with $0\leq \lambda \leq 1$ can be useful when there is a restricted access
to the state space of a given quantum system. Nevertheless, the accumulated
errors are scaled as $\left( 1-\lambda \right) ^{-1}$. One may observe such
a behavior also on the level of the reconstruction expression given by Eq.(%
\ref{rho_R}). In particular, the elements of the density matrix apparently
become singular when $\lambda \rightarrow 1$ since $\mu \sim (1-\lambda
)^{-1}$. Nevertheless, in the case of perfect measurements all such
\textquotedblleft singularities\textquotedblright\ automatically disappear.
Really, it can be easily seen from Eqs. (\ref{psi_s}), (\ref{psi_00})-(\ref%
{psi_0n}) that in this limit,
\begin{equation}
p_{sn}=\lambda p_{x0}+\sqrt{1-\lambda }\alpha _{sn}+O(1-\lambda ),
\label{psn_A}
\end{equation}%
where the functional dependence of $\alpha _{sn}$ on their indices is such
that by substituting Eq. (\ref{psn_A}) into Eq.(\ref{rho_R}) all singular
terms are canceled out.

It is worth noting that although the measurements are in the bases such that
all their elements are close the same state ($\sim \left\vert \psi
_{0}^{x}\right\rangle $), the reconstruction is done in the bi-orthogonal
bases. The separation between the elements of the bi-orthogonal bases, $\eta
$ tends to $-1/(p-1)$ as $\lambda \rightarrow 1$, which geometrically
corresponds to a homogenous division of the whole Hilbert space.

Of course, the presence even of small errors in measured probabilities $%
p_{sn}$ leads to substantial errors in the reconstructed density matrix Eq.(%
\ref{rho_R}) when the separation distance between elements of the bases
becomes smaller and smaller. In some sense this is the price to pay for
measuring in only a small part of the corresponding Hilbert space.

It is worth noting that in the particular case of two-dimensional systems, $%
p=2$, a different analysis can be done, see Appendix. In particular, two
non-orthogonal unbiased bases can be found such\ a way that these bases are
suffcient to resolve the identity and thus, form a POVM. As a result, the
same set of non-orthogonal bases can be used both to expand the density
matrix and to perform complementary measurements. Such approach is similar
to that developed in \cite{non ort tom} and\ closely related to the concept
of SIC-POVMs \cite{sic}.

In this article, we generalize the concept of mutually unbiased bases to include equally separated non-orthogonal bases.
Using such bases we have obtained an explicit expression for the optimal quantum
tomography in the non-orthogonal bases.

This work is partially supported by the Grants: 106525 of Consejo Nacional
de Ciencia y Tecnologia (CONACyT), Basal PFB0824, Milenio ICM P06-067F and
FONDECyT 1080535.

\section{Appendix: Non-orthogonal bases in dimension two}

In this Appendix we study the particular case of two-dimensional systems.

Let us consider a non-orthogonal basis in a two-dimensional Hilbert space,
which can be always expanded in an orthogonal basis $\{|j\rangle
,\left\langle i|j\right\rangle =\delta _{ij}$, $j,i=0,1\}$ as follows:
\begin{equation}
\left\vert \xi _{0}^{1}\right\rangle =\left\vert 0\right\rangle ,\quad
\left\vert \xi _{1}^{1}\right\rangle =e^{i\phi }\cos \theta \left\vert
0\right\rangle +\sin \theta \left\vert 1\right\rangle ,  \label{basis1}
\end{equation}
so that $\left\langle \xi _{0}^{1}|\xi _{1}^{1}\right\rangle =e^{i\phi
}\lambda _{1},\lambda _{1}=\cos \theta $, $0<\theta <\pi $ (observe, that here $\lambda$ is complex). It is well known
that in the orthogonal case, corresponding to $\theta =\pi /2$ there are two
orthonormal bases unbiased with (\ref{basis1}). Since any normalized basis
in a two dimensional Hilbert space can be represented as,
\begin{eqnarray*}
\left\vert \xi _{0}^{2}\right\rangle &=&e^{i\phi _{1}}\cos \theta
_{1}\left\vert 0\right\rangle +\sin \theta _{1}\left\vert 1\right\rangle , \\
\left\vert \xi _{1}^{2}\right\rangle &=&e^{i\phi _{2}}\cos \theta
_{2}\left\vert 0\right\rangle +\sin \theta _{2}\left\vert 1\right\rangle ,
\end{eqnarray*}%
the unbiasedement condition with the basis (\ref{basis1}),
\begin{equation}
|\langle \xi _{n}^{2}|\xi _{m}^{1}\rangle |=r,\quad \text{for all }n,m=0,1,
\end{equation}%
immediately leads to $\cos \theta _{1}=\cos \theta _{2}=r$ and to the
following restriction on the angles:
\begin{equation}
\cot 2\theta _{1}=\cot \theta \cos \left( \phi -\phi _{j}\right) ,\quad
~j=1,2.  \label{MUBCond}
\end{equation}%
In the orthogonal case, $\theta =\pi /2$, the only consequence of (\ref%
{MUBCond}) is that $\theta _{1}=\pi /4$ and thus $r=1/\sqrt{2}$, so that we
do not obtain any relation between the phases $\phi _{1,2}$. This allows to
find three unbiased basis that are well known. In the non-orthogonal case
the additional relation between the phases (\ref{MUBCond}) leads to the only
non-trivial solution;
\begin{equation*}
\phi _{2}=2\phi -\phi _{1},\quad \cos \left( \phi -\phi _{1}\right) =\frac{%
2r^{2}-1}{2r\lambda _{1}}\sqrt{\frac{1-\lambda _{1}^{2}}{1-r^{2}}},
\end{equation*}%
giving
\begin{eqnarray}
\left\vert \xi _{0}^{2}\right\rangle &=&e^{i\phi _{1}}\cos \theta
_{1}\left\vert 0\right\rangle +\sin \theta _{1}\left\vert 1\right\rangle ,
\label{basis2NOrt} \\
\quad \left\vert \xi _{1}^{2}\right\rangle &=&e^{i(2\phi -\phi _{1})}\cos
\theta _{1}\left\vert 0\right\rangle +\sin \theta _{1}\left\vert
1\right\rangle .  \notag
\end{eqnarray}%
This immediately means that there is no third basis simultaneously unbiased
with (\ref{basis1}) and (\ref{basis2NOrt}) in the non-orthogonal case. It is
worth noticing that these bases are different from the bases studied above,
in the sense that if the first basis (\ref{basis1}) is composed by the
eigenstates of the $Z$ operator, then the elements of the second basis (\ref%
{basis2NOrt}) are eigenstates of the operator $(ZX)^{\dag }$ when $\lambda_1$ is real. In other
words, if $\left\vert \xi _{n}^{1}\right\rangle =\left\vert \psi
_{n}^{0}\right\rangle$, then $\left\vert \xi _{n}^{2}\right\rangle
=\left\vert \phi _{n}^{1}\right\rangle $ for $n=0,1$.

Taking into account that the square overlap between elements of the second
basis is $|\left\langle \xi _{0}^{2}|\xi _{1}^{2}\right\rangle |^{2}=\lambda
_{2}^{2}=1-\sin ^{2}2\theta _{1}\sin ^{2}\left( \phi -\phi _{1}\right) $, we
obtain from (\ref{MUBCond}) the following relation between the possible
parameters of the bases
\begin{equation}
\lambda _{1}^{2}\lambda _{2}^{2}=(2r^{2}-1)^{2}.  \label{MUBNOrt}
\end{equation}
\textbf{The reconstruction relation}

In this case, there is no need to introduce an extra (orthogonal basis),
since the projectors $\left\vert \xi _{j}^{s}\right\rangle \left\langle \xi
_{j}^{s}\right\vert $ resolve the identity
\begin{equation}
\sum_{s=1}^{2}\sum_{j=0}^{1}\left\vert \xi _{j}^{s}\right\rangle
\left\langle \xi _{j}^{s}\right\vert =2\hat{\mathbf{I}},  \label{identity}
\end{equation}%
if the relation between $\lambda _{1}$ and $\lambda _{2}$,
\begin{equation}
~\lambda _{1}^{2}=\lambda _{2}^{2}=1-2r^{2}  \label{restr_lr}
\end{equation}%
is satisfied. Since $\lambda _{1,2}$ are real, the relation (\ref{restr_lr})
automatically imposes the restriction $r\leq 1/\sqrt{2}$ over the possible
values of $r$.

The above expression allows us to reconstruct the density matrix
\begin{equation}
\hat{\rho}=\sum_{s=1}^{2}\sum_{j=0}^{1}q_{sj}\left\vert \xi
_{j}^{s}\right\rangle \left\langle \xi _{j}^{s}\right\vert ,  \label{density}
\end{equation}%
in terms of measured probabilities $p_{sj}=\text{Tr}(\hat{\rho}\left\vert
\xi _{j}^{s}\right\rangle \left\langle \xi _{j}^{s}\right\vert
)=\left\langle \xi _{j}^{s}\right\vert \hat{\rho}\left\vert \xi
_{j}^{s}\right\rangle $, where $s=1,2,j=0,1,$ giving
\begin{eqnarray}
q_{s0} &=&\frac{(1-r^{2})p_{s0}-(1-3r^{2})p_{s1}-2r^{4}}{2r^{2}(1-2r^{2})},
\notag \\
q_{s1} &=&\frac{(1-r^{2})p_{s1}-(1-3r^{2})p_{s0}-2r^{4}}{2r^{2}(1-2r^{2})},
\label{qkj}
\end{eqnarray}%
where $0<r^{2}<1/2$. The minimum error corresponds to the case when each
measurement gives a single element of the density matrix in the
representation (\ref{density}), i.e. when $\lambda _{1}^{2}=\lambda
_{2}^{2}=r^{2}=1/3$, this case corresponds to SIC-POVM \cite{sic},
\begin{equation}
\hat{\rho}=3\sum_{\alpha =1}^{2}\sum_{j=0}^{1}p_{\alpha j}\left\vert \psi
_{j}^{\alpha }\right\rangle \left\langle \psi _{j}^{\alpha }\right\vert -%
\hat{1}.  \notag
\end{equation}

\end{document}